%% file: v2.tex
\documentclass[structabstract]{aa}  
\usepackage{graphicx}
\usepackage{txfonts}

\usepackage{natbib}
\bibpunct[, ]{(}{)}{;}{a}{}{,}
\newcommand{\griz}{$g^\prime r^\prime i^\prime z^\prime$~}
\newcommand{\JHK}{$JHK_S$~}
\newcommand{\gK}{$g^\prime r^\prime i^\prime z^\prime JHK_S$~}
\begin{document}
\title{The bright optical/NIR afterglow of the faint GRB~080710 - Evidence for a jet viewed off axis}


\author{T. Kr\"{u}hler \inst{1,2}
 		  \and
          J. Greiner \inst{1}
          \and      
          P. Afonso \inst{1}
          \and          
          D. Burlon \inst{1}     
          \and 
          C. Clemens \inst{1}
		  \and
          R. Filgas \inst{1}
          \and
          D. A. Kann \inst{3}
          \and
          S. Klose \inst{3}
          \and
          A. K\"{u}pc\"{u} Yolda\c{s} \inst{5}
          \and
          S. McBreen \inst{4}
          \and
          F. Olivares \inst{1}
          \and
          A. Rau \inst{1}
          \and
          A. Rossi \inst{3}
          \and
          S. Schulze \inst{3,6}
          \and
          G. P. Szokoly \inst{7}
          \and
          A. Updike \inst{8}
          \and
          A. Yolda\c{s} \inst{1}
}

\institute{Max-Planck-Institut f\"{u}r extraterrestrische Physik, Giessenbachstrasse, 85748 Garching, Germany.
            \email{kruehler@mpe.mpg.de}
            \and
            Universe Cluster, Technische Universit\"{a}t M\"{u}nchen, Boltzmannstrasse 2, 85748 Garching, Germany.
            \and
           Th\"{u}ringer Landessternwarte Tautenburg, Sternwarte 5, 07778 Tautenburg, Germany
           \and
           School of Physics, University College Dublin, Dublin 4, Ireland
            \and
           European Southern Observatory, 85748 Garching, Germany.
           \and
           Center for Astrophysics and Cosmology, University of Iceland, Dunhagi 5, 107 Reykjav\'ik, Iceland
           \and
           Institute of Physics, E\"{o}tv\"{o}s University, P\'{a}zm\'{a}ny P. s. 1/A, 1117 Budapest, Hungary
           \and
           Department of Physics and Astronomy, Clemson University, Clemson, SC 29634, USA.
            }

\date{} 

 
\abstract
{}
{We investigate the optical/near-infrared light curve of the afterglow of GRB~080710 in the context of rising afterglows.}
{Optical and near-infrared photometry was performed using the seven channel imager GROND and the Tautenburg Schmidt telescope. X-ray data were provided by the X-ray Telescope onboard the \textit{Swift} satellite. We construct an empirical light curve model using the available broadband data, which is well-sampled in the time and frequency domains.}
{The optical/NIR light curve of the afterglow of GRB~080710 is dominated by an initial increase in brightness, which smoothly turns over into a shallow power law decay. At around 10~ks post burst, there is an achromatic break from shallow to steep decline in the afterglow light curve with a change in the power law index of $\Delta\alpha\sim$0.9.}
{The initially rising achromatic light curve of the afterglow of GRB 080710 can be accounted for with a model of a burst viewed off-axis or a single jet in its pre deceleration phase and in an on-axis geometry. An unified picture of the afterglow light curve and prompt emission properties can be obtained with an off-axis geometry, suggesting that late and shallow rising optical light curves of GRB afterglows might be produced by geometric effects.}


\maketitle
%

\section{Introduction}

The launch of the \textit{Swift} satellite \citep{geh04} in 2004 opened a new field of Gamma-Ray Burst (GRB) afterglow physics. With its precise localization by the Burst Alert Telescope (BAT; \citealp{bar05}), rapid slewing capabilities and early follow up with two instruments in the X-ray and ultraviolet/optical regime, studies of the early afterglow phase were possible for the first time with larger sample statistics of around 50 per year.

Long GRBs are generally classified according to the spectral properties of their prompt emission. While conventional GRBs (CGRBs) have the peak energy of their observed spectrum in the 300~keV range \citep{pre00}, the spectrum of X-ray rich bursts (XRRs) and X-ray flashes (XRFs) peak at significantly lower energies, typically around 50~keV for XRRs or 10~keV for XRFs respectively \citep[e.g.][]{hei01, kip03}. The spectral and temporal properties \citep[e.g.][]{sak05} and their similar afterglows as compared to CGRBs provide strong evidence, though, that XRRs/XRFs represent a softer region of a continuous GRB distribution \citep[e.g.][]{lam05, sak08}.

An unified picture of the subclasses of GRBs can be obtained when attributing the observed differences in their peak energy to different angles with respect to the symmetry axis of the jet \citep[e.g.][]{yam02}. The kinetic energy in the jet per solid angle $\varepsilon$ is usually parametrized as a top hat \citep[e.g.][]{rho99, woo99}, Gaussian \citep{zha02b}, power law structured outflow with $\varepsilon\propto(\theta/\theta_{jet})^{-q}$ \citep{mes98}, or a top hat with lower energetic wings. The resulting shape of the afterglow light curve then depends on the viewing angle and jet structure \citep[e.g.][]{ros02}. 

In an inhomogeneous jet model, the initial bulk Lorentz factor as well as the specific deceleration time and radius are dependent on the distance to the symmetry axis of the jet \citep{kum03}. Hence, a geometric offset of the observers line of sight from the jets symmetry axis will have a distinct signature in the optical light curve \citep[e.g.][]{gra03a}. Due to the relativistic beaming of the decelerating ejecta, an observer located off-axis to the central jet will see a rising optical afterglow light curve at early times \citep[e.g.][]{pan98, gra02}. The steepness of the rise would then be characteristic of the off-axis angle and the jet structure: the farther the observer is located from the central emitting cone or the faster the energy per solid angle decreases outside the jet, the shallower is the observed rise in a structured jet model \citep{pan08}. A restframe peak energy E$^{\rm rest}_{\rm peak}$ consistent with an XRF would thus correspond to a shallow rise or early plateau phase of the afterglow. With decreasing off-axis angle, both E$^{\rm rest}_{\rm peak}$ and the optical afterglow rise index will increase to XRRs and steeper rising early afterglow light curves.
 
\section{Observations}

At T$_{\rm 0}$=07:13:10 UT on 10 July 2008, \textit{Swift} triggered and located GRB~080710, but did not slew immediately to the burst \citep{sba08a}. Due to an observing constraint, observations with the two narrow field instruments, the X-ray-(XRT; \citealp{bur05}) and UV/Optical Telescope (UVOT; \citealp{rom05}) started 0.87~h and 0.89~h after the trigger \citep{lan08}. The burst had a relatively smooth fast rise - exponential decay structure with T$_{\rm 90}$(15-350~keV) = 120$\pm$17~s and weak indication of a precursor 120~s before the main peak \citep{tue08}. Above 100~keV, the burst is only marginally detected by BAT and its spectrum is well described with a single power law of index $-$1.47$\pm$0.23 with a total fluence in the 15-150~keV range of 1.4$\pm$0.2$\cdot$10$^{-6}$ erg/cm$^2$ \citep{tue08}. Using the spectral slope from the BAT data, and following \citet{sak09}, the peak energy of the prompt emission spectrum can be constrained to 110$^{+340}_{-60}$~keV including the uncertainties of the BAT power law slope. The fluence ratio of GRB~080710 between the two BAT bands 25-50~keV and 50-100~keV is S(25-50~keV)/S(50-100~keV) = 0.70$\pm$0.15, and the burst thus qualifies as a CGRB in the observers frame, with errors ranging into XRRs when applying the working definition of \citet{sak08}.

Assuming a spectral shape of a Band function \citep{ban93} with a peak energy of around 110~keV and a high energy index of $-$2.5, standard $\Lambda$CDM cosmology ($\Omega_M$=0.27, $\Omega_{\Lambda}$=0.73, H$_0$=71(km/s)/Mpc) and a redshift of 0.845 \citep{per08b, fyn09} we derive a bolometric energy release for GRB~080710 of log~E$_{\gamma \rm ,iso}$[erg]=51.75 with a restframe peak energy of E$^{\rm rest}_{\rm peak}\sim$200~keV. 
Peak energies of the observed prompt spectrum of 50~keV, 300~keV or 500~keV result in log~E$_{\gamma \rm ,iso}$[erg]$\approx$51.70, 51.94 or 52.14, respectively. Compared to a sample of previous bursts with known redshift \citep[e.g.][]{ama08}, these estimates put GRB~080710 to the lower energy end of GRBs, with an inferred bolometric energy release of around 10$^3$ times less than the extremely energetic GRB~080916C \citep{abd09, gre09}. Hence, a low E$^{\rm rest}_{\rm peak}$ in the 50-200~keV range is also supported by the Amati relation \citep{ama02}, consistent with the best estimate value using the BAT spectral slope. Given the low redshift and the prompt emission properties for GRB~080710, it seems thus very likely that E$^{\rm rest}_{\rm peak}$ is in a range which is typically associated with a XRR in the GRB rest frame (100-300~keV, \citealp{sak08}), though a hard burst cannot be completely ruled out by the observations.

GROND \citep{gre08} at the 2.2~m MPI/ESO telescope at LaSilla observatory responded to the \textit{Swift} trigger and initiated automated observations which started 384~s after the burst. During the first two hours only the \griz CCDs of GROND were operating. Observations in all seven colors \gK simultaneously started 1.98~h later and continued until the start of the local nautical twilight at 10:27 UT. Afterwards, GROND switched to a NIR-only mode, where only imaging in \JHK was performed. TLS imaging was obtained between 00:09 UT and 01:43 UT on 11 July 2008 in filters $BVR$ and $I$ \citep{sch08}. In addition, GROND imaged the field of GRB~080710 3 and 4 days after the burst.

The XRT light curve was downloaded from the XRT light curve repository \citep{eva07} and spectra were obtained with the \texttt{xrtpipeline} tool using the latest calibration frames from the \textit{Swift} CALDB and standard parameters. The spectra were fitted using the XSPEC package \citep{arn96} with a foreground hydrogen column density at the Galactic value of $N_{\rm H}$=4.1$\times 10^{20}$~cm$^{-2}$ \citep{kal05}. Optical/NIR data (see Tab.~\ref{grizphot} and \ref{JHKphot}) were reduced using standard IRAF tasks \citep{tod93} similar to the procedure outlined in \citet{tk08}.

\section{Results}

\subsection{Afterglow light curve}

\begin{figure}
\centering
\includegraphics[width=\columnwidth]{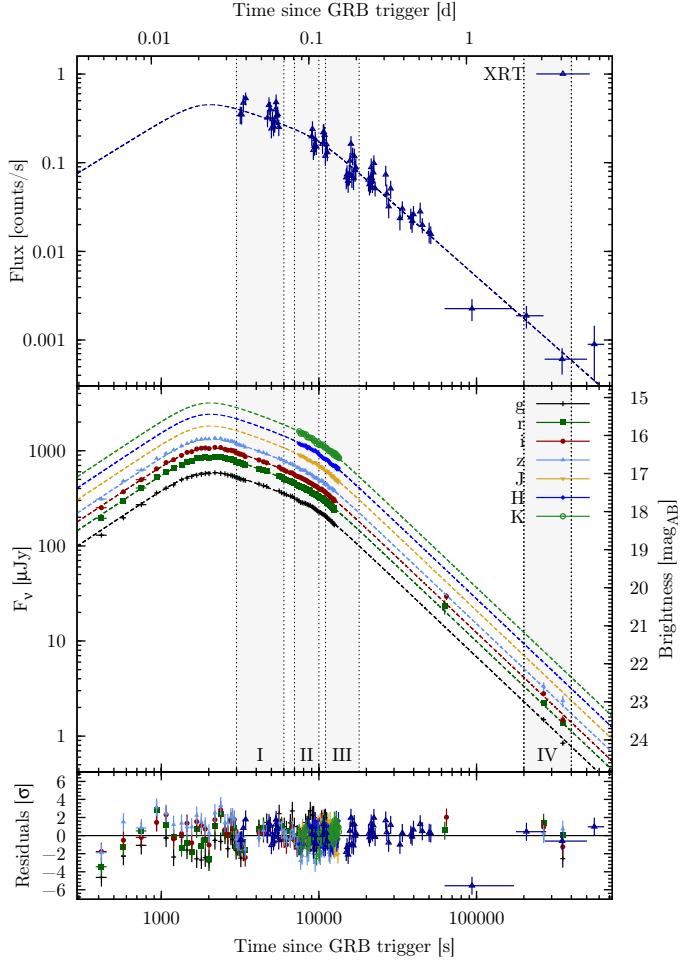}
\caption{Light curves of the X-ray (top panel) and optical/NIR (middle panel) afterglow of GRB~080710. Residuals to the combined light curve fit are shown in the lowest panel. Shown data are not corrected for Galactic foreground reddening. Upper limits are not shown to enhance clarity.}
\label{LC}
\end{figure}

The optical light curve (Fig.~\ref{LC}) exhibits two salient features during the observation. First, it shows an initial rise in brightness up to a peak at around 2000~s, and second, there is a break in the light curve at roughly 10~ks. 

The light curve was parametrized with an empirical model of three smoothly connected power laws. The global $\chi^2$ of $F_{\nu, i}(t)$, where $i$ denotes the individual filter or bandpass, was minimized by assuming an achromatic functional form of $F_{\nu, i}(t)$=$\eta_{\nu, i} \times F_{\nu}(t)$ where only the overall flux normalization $\eta_{\nu, i}$ depends on the filter. $F_{\nu}(t)$ was adapted from \citet{lia08}. 
As a result of the high precision of the data and good sampling in time domain, all parameters were left free to vary and are presented in Tab.~\ref{LCfits}. In principle, all fit parameters depend on the choice of T$_{0}$. Setting T$_{0}$ to the time of the precursor (i.e. -120~s), we find that the fit parameters describing the early/late power laws vary by a maximum of 20\% and 2\%, respectively. Hence, the uncertainty in T$_{0}$ does not change the obtained results significantly or affect the overall conclusions.

Given that the decay after the peak at 2~ks with an index of $-$0.63$\pm$0.02 is too shallow to be explained as the normal decay phase and the late temporal slope of $-$1.57$\pm$0.01 is roughly consistent with the closure relations for the normal decay in the $\nu_m < \nu < \nu_c$ regime 
for a homogeneous ISM and slow cooling case ($\alpha$=3$\beta$/2), there is no apparent evidence for a jet-break before 350~ks, and thus $\theta_{\rm jet} > 10^\circ$ according to \citet{sar99}.  

\subsection{Broad-band spectrum}

\begin{figure}
\centering
\includegraphics[angle=270,width=\columnwidth]{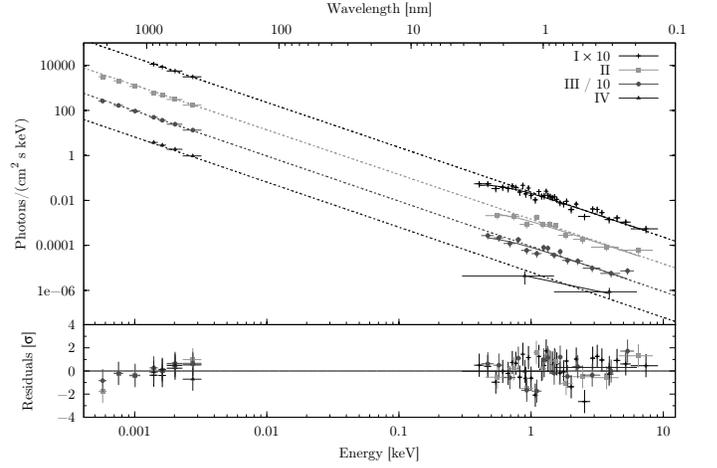}
\caption{Broad band spectral energy distribution from XRT and GROND at different epochs (upper panel). The data were fitted with a power-law, modified by a Galactic and intrinsic hydrogen column. The best fit power law is shown in dotted lines, the best fit model including the soft X-ray absorption in solid lines. In the lower panel the residuals of the data to the best fit model.}
\label{SED}
\end{figure}

Using the optical/NIR and X-ray data, the afterglow spectrum can be constrained over a broad wavelength range. Four different time intervals were selected to construct a broad band spectral energy distribution (SED, Fig.~\ref{SED}). The different epochs are indicated in the light curve plot with shaded regions, and the SED fit parameters are presented in Tab.~\ref{SEDfits}.

As already indicated by the light curve, there is no sign of spectral evolution throughout the observation. Both the early turnover from rising to falling, as well as the second break are achromatic with high measurement accuracy. The optical/NIR SED alone is consistent with a power law of the X-ray spectral index without strong signatures of curvature due to intrinsic reddening. The expected Galactic foreground extinction $A_{\rm V}$=0.23~mag \citep{schl98} however is significant, so some amount of host extinction might be masked by the uncertainty in the foreground correction. In addition, the obtained optical data hardly probe the rest frame UV regime, where most of any intrinsic extinction would be apparent. 

Given that the light curve evolution is similar in both energy ranges and the extrapolation of the X-ray data nicely matches the optical flux, i.e. $\beta_{\rm opt}\sim\beta_{\rm ox}\sim\beta_{\rm x}$, both the optical/NIR and X-ray emission probe the same segment of the afterglow synchrotron spectrum. This implies that the X-ray and optical data are above the typical synchrotron frequency $\nu_{m}$ and in the spectral regime of max($\nu_m, \nu_c)<\nu_{\rm opt}<\nu_X$, or $\nu_m<\nu_{\rm opt}<\nu_X<\nu_c$, where the latter is consistent with fireball model in a homogeneous ISM and slow cooling case. The spectral index of the electron distribution $p$ would then be $p$=2$\beta$=2.00$\pm$0.02 or 2$\beta$+1=3.00$\pm$0.02, respectively. Given that not all bursts are consistent with the closure relations in the basic fireball scenario \citep[e.g.][]{eva09}, we consider both cases in the following. Consequentially, the expected break in the synchrotron afterglow spectrum at the cooling frequency $\nu_c$ could be below the optical at the start of the observations 6 minutes after the burst, or, assuming $\nu_m<\nu<\nu_c$, above the X-rays for the whole observational period.

\section{Discussion}

A number of previous bursts have shown a rising optical afterglow at early times, e.g. GRBs~060418, 060607A \citep{mol07} amongst others \citep[e.g.][]{tk08, fer09, gre09a, oat09, ryk09, klo09}. Similar to the X-Ray Flash 071031 \citep{tk09}, the optical SED does not show significant evolution during the rise, and all bands peak at the same time. 

An achromatic rising component is generally related to either the onset of the fireball forward shock \citep[e.g.][]{sar99} seen face-on, or to an outflow seen off-axis \citep[e.g.][]{pan98}. In the first case, the apparent increase in brightness is caused by the increasing number of radiating electrons. The time of the light curve peak at $T_0$+2~ks is much later than the end of significant $\gamma$ emission ($T_0$+40~s), so the afterglow can be described in the thin shell approximation. The jet is then expected to produce a peak in the light curve when the swept up circumburst medium efficiently decelerates the ejecta. Depending on the profile of the circumburst medium, the rise has indices of $\sim$2 ($\nu_{\rm c}<\nu_{\rm opt}$) or 3 ($\nu_{\rm c}>\nu_{\rm opt}$) in an ISM, or $\sim$0.5 in an wind environment \citep{pan08}. Given that the majority of bursts prefer an ISM profile, and the late afterglow decline is consistent with it, 
we thus consider only the ISM thin shell case in the following.  

In the off-axis case the peak is a geometric effect: as the shock wave decelerates, the relativistically beamed emission cone widens and gradually enters the sight line of the observer. The light curve morphology is then dependent on the jet's structure and off axis angle $\theta_{obs}$, and reaches a maximum when $\Gamma \sim (\theta_{\rm obs}- \theta_{\rm c})^{-1}$ where $\theta_{\rm c}$ is the angle of an uniform cone around the symmetry axis of the jet.

There is no evidence of chromatic evolution, which would be the case if the peak was caused by a moving $\nu_m$ through the optical bands or dust destruction, and none of these processes produce the early rise. In addition, there is also no sign of a reverse shock, which is expected to decline with a temporal index of $-$1.75 for $p$=2 or $-$2.5 for $p$=3. The latter, however, might be masked by a dominating forward shock, or happened before the start of the GROND observations.

\subsection{Decelerating ejecta in an on-axis geometry}

If the light curve peak was caused by a jet in its pre deceleration phase, conclusions about the motion of the ultra-relativistic outflow from the central engine can be drawn. Using the time of the absolute light curve maximum $t_{\rm max}\approx$ 2~ks, log~E$_{\gamma \rm ,iso}$[erg]=51.70-52.14 and following \citet{mol07}, we find initial Lorentz factors of the bulk outflow of around $\Gamma^{\rm ISM}_0\approx$ 90-100 ($\Gamma^{\rm wind}_0\approx$30-40). This is at the very low end of the theoretically expected velocity of the outflow to produce $\gamma$-rays \citep[e.g.][]{pir05}, and together with the divergence in the measured (1.1) and expected ($\sim$2-3) rise index, makes the scenario of a single on-axis decelerating jet somewhat contrived. In addition, there seems to be a small population of late-peaking afterglows or long plateaus (e.g. XRF~030723 \citep{fyn04} or GRB~060614 \citep{del06}) where the derived Lorentz-factor in an on-axis geometry from the optical afterglow peak are uncomfortably small. Furthermore, all previously observed rise indices have a broad distribution (cp. e.g.~\citet[][]{pan08, oat09, ryk09, klo09} and references therein) from early plateaus to very fast rising curves, and they do not cluster around the expected $t^{2-3}$. Consequentially, it seems plausible that at least some rising afterglows are not caused by the onset of the afterglow, but rather by a geometrical offset of the observers sight line with respect to the jets central cone.

\subsection{Jet seen off-axis}

Contrary to the model of an on-axis jet in its pre deceleration phase, an off-axis scenario is able to account for a broad range of observed rise indices. The peak time and rise index then relates to the off-axis angle or jet structure and therefore could describe a large diversity of early afterglows in a single framework \citep{pan08}.

If the energy in the jet outer wings decreases rapidly, the early emission of the line of sight ejecta is negligible as compared to the central part, and the jet structure can be approximated as a homogeneous top-hat, where the burst energetics can be used to constrain the offset angle. Following \citet{gra02}, a homogeneous jet with a half opening angle $\theta_{\rm jet}$ and a Lorentz factor $\Gamma$ seen off-axis at an angle $\theta_{\rm obs}$ will appear less energetic by a factor of $b^6$, where $b=\Gamma\, (\theta_{\rm obs} - \theta_{\rm jet})$. Assuming a mean value of $\log E_{\gamma ,\rm iso}$ [erg] = 53 and, hence adopting $b^6\lesssim$10 for GRB 080710, it follows $\theta_{\rm obs} - \theta_{\rm jet} \lesssim 3^\circ/\Gamma_{\rm 100}$. E$_{\rm peak}$, if viewed on-axis, would then be $b^2\,$E$_{\rm peak}^{\rm obs}\approx$~300~keV.

However, the jet geometry does not necessarily have to be a simple top-hat. In a realistic jet model, the jet viewed off-axis is inhomogeneous, i.e. has a top-hat structure with wings of lower energy, or is Gaussian shaped \citep[e.g.][]{zhang03, eic06}. In addition, some bursts show evidence that their jet structure is a composition of two jets \citep{ber03, gran06,rac08}. In this two component jet model a narrow, fast jet produces the $\gamma$-rays and early afterglow, and a slow wide jet dominates the late afterglow emission \citep{pen05}.

\begin{figure}
\centering
\includegraphics[angle=270,width=\columnwidth]{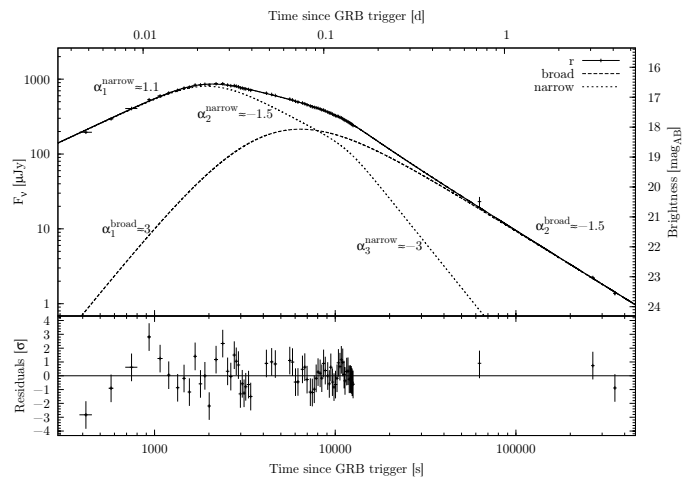}
\caption{Two component fit for GRB~080710 as the superposition of the afterglow of two jets with $\nu_m<\nu_{\rm opt}<\nu_X<\nu_c$ and p$\sim$3 for both components. The narrow jet is viewed slightly off-axis and produces a shallow rise as its emission spreads during deceleration due to relativistic beaming effects. The broad jet is viewed on-axis with $\Gamma_0\sim 50$, $\theta_{\rm jet} > 10^\circ$ and has the expected steep rise during its pre deceleration phase. Shown is the GROND $r^\prime$ band data, all other bands are omitted to enhance clarity.}
\label{2LC}
\end{figure}

In these cases, the resulting afterglow light curve in an off-axis geometry is a superposition of two different components: the emission from the ejecta with lower Lorentz-factors, typically dominating at late times, and the relativistic spreading of the decelerating jet around the symmetry axis. The relative energies, jet structure and offset angle then define the light curve morphology. In particular, the delayed onset of the broad jet emission in its pre deceleration phase might be responsible for the shallow decay observed after the peak. Remarkably, the light curve is equally well ($\chi^2$=485 for 425 d.o.f) fit using the sum of the afterglow of two jets, where the narrow one is viewed slightly off-axis (Fig.~\ref{2LC}). Hence, the shallow decay phase could be the result of the superposition of the normal decay phase of the narrow-jet afterglow and the rise of the broad jet with $\Gamma_0\sim$50, $\theta_{\rm jet} > 10^\circ$ in its pre deceleration phase. After the emergence of the broad jet afterglow, it subsequently dominates the light curve morphology (Fig.~\ref{2LC}). The two component model thus provides a phenomenological explanation for the shallow decay phase by attributing the shallow slope to the increasing energy dissipation in the pre deceleration phase of the broader jet in a specific jet configuration. The opening angle of the narrow jet can be constrained from the light curve fitting to around 2-4$^\circ$, but its jet break is masked by the brighter broad jet at later times (Fig.~\ref{2LC}). An alternative, jet geometry independent mechanism of energy injection during the shallow decay phase is the refreshed shock scenario \citep[e.g.][]{ree98, zha06}. A long lived central engine or a simultaneous ejection of shells with a distribution of Lorentz factors could cause the continuous energy injection required for a shallow decay \citep[e.g.][]{nou06}.

An off-axis viewing angle in a two component or structured jet model with an energy injection can thus provide a consistent picture for the light curve morphology and the relatively low estimates of E$_{\gamma,~\rm iso}$ and E$_{\rm peak}^{\rm rest}$ of the prompt emission of GRB~080710. In an off-axis scenario, a lower E$_{\rm peak}^{\rm rest}$ of the prompt emission spectrum would correspond to a later and fainter afterglow maximum, as both are caused by geometric effects. We caution that the spectral properties of BAT bursts are generally not well constrained, and GRB~080710 is no exception in this aspect. The BAT data, however, indicate a mildly soft event, which could be associated with a XRR in the bursts restframe, consistent with the off-axis interpretation of the optical light curve in an unified model.  

\section{Conclusions}

The broad-band light curve of the afterglow of GRB~080710 shows two salient features, both achromatic with high precision: an early rise in its brightness, peaking at $\sim$2~ks, and a turnover from a shallow to steep decline at $\sim$10~ks. The early rise can be caused by a jet in its pre deceleration phase, or a viewing angle outside the central cone. The latter scenario is naturally able to explain a late-rising afterglow for a soft and weak burst due to a viewing angle offset with respect to the symmetry axis of the jet. An off-axis scenario provides a consistent description of the properties of GRB~080710, and can additionally account for a broad range of rise indices. Consequentially, some of the rising afterglow light curves, especially late and shallow ones, might not represent the same class of afterglows which rise due to increasing emission in the pre deceleration phase, but rather provide evidence for an off-axis location of the observer.
The achromatic early increase in brightness observed in the mildly soft GRB~080710 is too shallow to be accounted for with the onset of the afterglow, but significantly steeper than recently observed in the XRFs 071031 \citep{tk09} and 080330 \citep{gui09}. This might already hint on a common dependence of E$^{\rm rest}_{\rm peak}$ and the rise index of the early optical light curve on the off-axis angle as expected in an unified model: the softer the prompt emission, the more off-axis, and thus the shallower the rise. It remains to be tested with a larger sample of early afterglows with well constrained energetics and light curves of the prompt emission from combined \textit{Swift}/BAT and \textit{Fermi}/GBM detections, whether and how the structure of an early rise in the optical afterglow is related to prompt emission properties, and in particular, the rest frame E$^{\rm rest}_{\rm peak}$ and E$_{\gamma, \rm iso}$. A possible correlation would then shed light on the nature of the early afterglow rise, the shallow decay segment, and the jet structure in general.
 
\begin{acknowledgements}
We thank the referee for very helpful comments, which helped to increase the quality of the paper significantly. TK acknowledges support by the DFG cluster of excellence 'Origin and Structure of the Universe'. A.R. and S.K. acknowledge support by DFG grant Kl 766/11-3. Part of the funding for GROND (both hardware and personnel) was granted from the Leibniz-Prize to Prof. G. Hasinger (DFG grant HA 1850/28-1). SS acknowledges support by a Grant of Excellence from the Icelandic Research Fund. This work made use of data supplied by the UK Swift Science Data Centre at the University of Leicester.

\end{acknowledgements}

\input{bib.tex}

\Online

\onlfig{1}{
\begin{figure}
\centering
\includegraphics[width=\columnwidth]{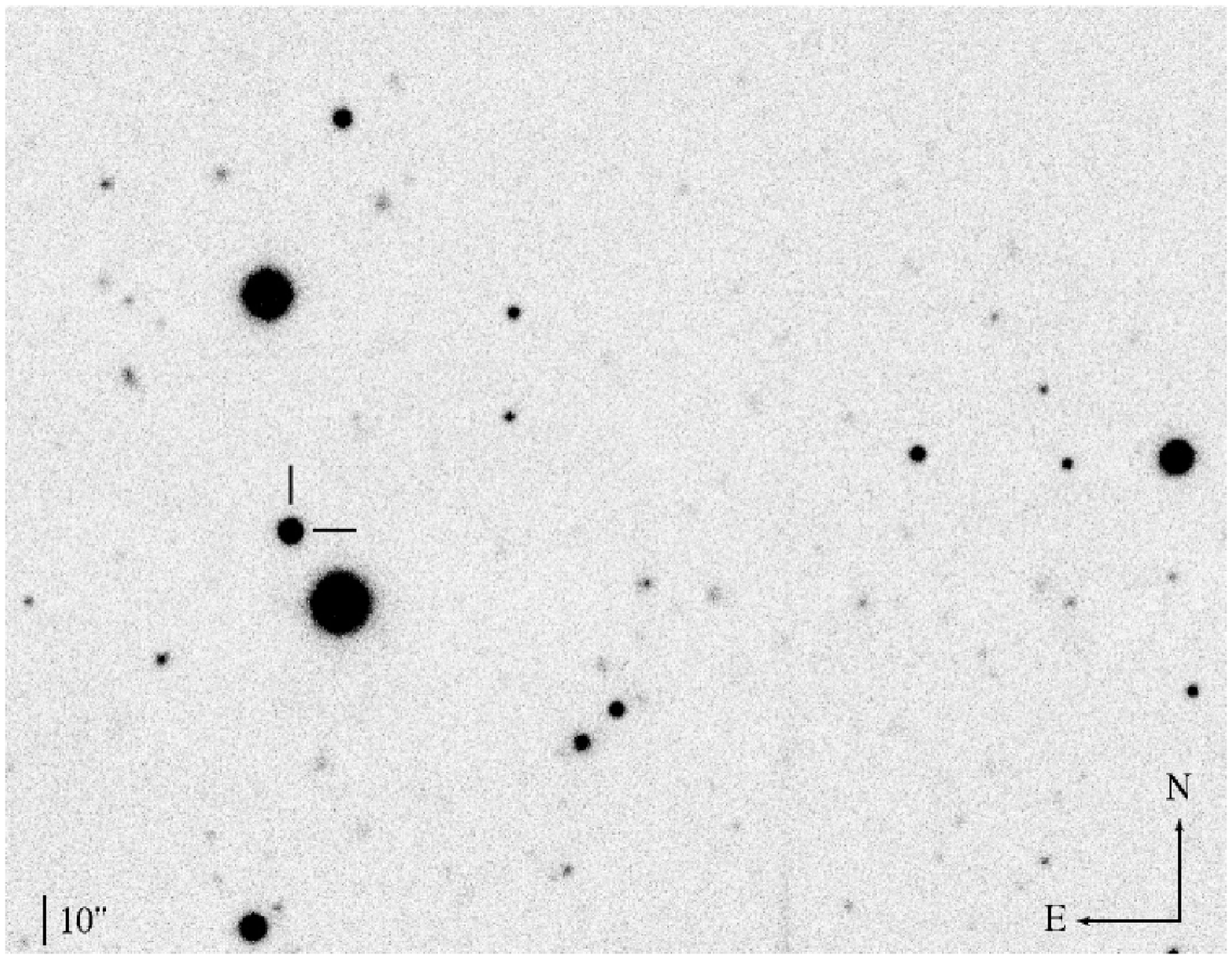}
\caption{GROND $r^\prime$ band image of the field of GRB~080710 obtained approximately 2~ks after T$_{0}$. The optical afterglow is marked and the shown image is roughly 4.2$^\prime$ by 3.2$^\prime$.}
\label{pic}
\end{figure}
}
\input{tab0.tex}
\input{tab01.tex}
\input{tab1.tex}
\input{tab2.tex}

\end{document}

%% file: bib.tex
\hyphenation{Post-Script Sprin-ger}

%% file: tab0.tex
\onltab{1}{
\begin{table*}
\caption{griz photometric data \label{grizphot}}
\begin{tabular}{ccccccc}
\hline
\noalign{\smallskip}
$T_{\rm mid} - T_{\rm 0} $ & Exposure & Filter & \multicolumn{4}{c}{Brightness$^{(a)}$}  \\  
$[\rm ks]$ & $[\rm s]$ &  & \multicolumn{4}{c}{mag$_{\rm AB}^{(bc)}$}  \\ 
\hline
 &  &  & $g^\prime$ & $r^\prime$ & $i^\prime$ & $z^\prime$ \\
\hline
0.4169 & 66 & \griz & 18.619 $\pm$ 0.028 & 18.175 $\pm$ 0.026 & 17.899 $\pm$ 0.026 & 17.673 $\pm$ 0.028 \\
0.5745 & 35 & \griz & 18.162 $\pm$ 0.024 & 17.724 $\pm$ 0.018 & 17.478 $\pm$ 0.018 & 17.200 $\pm$ 0.019 \\
0.7457 & 115 & \griz & 17.815 $\pm$ 0.016 & 17.381 $\pm$ 0.010 & 17.162 $\pm$ 0.015 & 16.909 $\pm$ 0.016 \\
0.9334 & 35 & \griz & 17.509 $\pm$ 0.013 & 17.086 $\pm$ 0.009 & 16.875 $\pm$ 0.011 & 16.604 $\pm$ 0.016 \\
1.0726 & 71 & \griz & 17.385 $\pm$ 0.008 & 16.961 $\pm$ 0.006 & 16.711 $\pm$ 0.009 & 16.474 $\pm$ 0.011 \\
1.1977 & 35 & \griz & 17.300 $\pm$ 0.008 & 16.865 $\pm$ 0.006 & 16.637 $\pm$ 0.008 & 16.381 $\pm$ 0.009 \\
1.3423 & 35 & \griz & 17.163 $\pm$ 0.007 & 16.773 $\pm$ 0.006 & 16.523 $\pm$ 0.007 & 16.274 $\pm$ 0.008 \\
1.4556 & 35 & \griz & 17.118 $\pm$ 0.007 & 16.701 $\pm$ 0.005 & 16.442 $\pm$ 0.007 & 16.215 $\pm$ 0.007 \\
1.5595 & 35 & \griz & 17.075 $\pm$ 0.007 & 16.668 $\pm$ 0.005 & 16.425 $\pm$ 0.007 & 16.169 $\pm$ 0.007 \\
1.6767 & 35 & \griz & 17.044 $\pm$ 0.006 & 16.596 $\pm$ 0.005 & 16.354 $\pm$ 0.007 & 16.112 $\pm$ 0.007 \\
1.7947 & 35 & \griz & 17.020 $\pm$ 0.006 & 16.595 $\pm$ 0.005 & 16.334 $\pm$ 0.007 & 16.095 $\pm$ 0.008 \\
1.8996 & 35 & \griz & 17.003 $\pm$ 0.006 & 16.573 $\pm$ 0.005 & 16.326 $\pm$ 0.007 & 16.089 $\pm$ 0.007 \\
2.0065 & 35 & \griz & 17.003 $\pm$ 0.006 & 16.593 $\pm$ 0.005 & 16.343 $\pm$ 0.006 & 16.094 $\pm$ 0.006 \\
2.1892 & 115 & \griz & 16.976 $\pm$ 0.005 & 16.552 $\pm$ 0.004 & 16.312 $\pm$ 0.004 & 16.070 $\pm$ 0.005 \\
2.3835 & 115 & \griz & 16.995 $\pm$ 0.005 & 16.553 $\pm$ 0.004 & 16.315 $\pm$ 0.004 & 16.078 $\pm$ 0.005 \\
2.5359 & 35 & \griz & 17.007 $\pm$ 0.006 & 16.597 $\pm$ 0.004 & 16.363 $\pm$ 0.007 & 16.121 $\pm$ 0.006 \\
2.6421 & 35 & \griz & 17.042 $\pm$ 0.006 & 16.617 $\pm$ 0.005 & 16.386 $\pm$ 0.007 & 16.136 $\pm$ 0.007 \\
2.7614 & 35 & \griz & 17.044 $\pm$ 0.006 & 16.617 $\pm$ 0.005 & 16.387 $\pm$ 0.006 & 16.147 $\pm$ 0.007 \\
2.8323 & 35 & \griz & 17.061 $\pm$ 0.006 & 16.635 $\pm$ 0.005 & 16.401 $\pm$ 0.007 & 16.159 $\pm$ 0.007 \\
2.9105 & 35 & \griz & 17.074 $\pm$ 0.006 & 16.652 $\pm$ 0.005 & 16.409 $\pm$ 0.007 & 16.176 $\pm$ 0.007 \\
2.9854 & 35 & \griz & 17.103 $\pm$ 0.006 & 16.692 $\pm$ 0.005 & 16.453 $\pm$ 0.007 & 16.219 $\pm$ 0.007 \\
3.0564 & 35 & \griz & 17.110 $\pm$ 0.006 & 16.696 $\pm$ 0.005 & 16.463 $\pm$ 0.007 & 16.236 $\pm$ 0.007 \\
3.1276 & 35 & \griz & 17.125 $\pm$ 0.007 & 16.718 $\pm$ 0.005 & 16.483 $\pm$ 0.007 & 16.248 $\pm$ 0.008 \\
3.1962 & 35 & \griz & 17.151 $\pm$ 0.006 & 16.724 $\pm$ 0.005 & 16.492 $\pm$ 0.007 & 16.261 $\pm$ 0.008 \\
3.3151 & 35 & \griz & 17.164 $\pm$ 0.006 & 16.744 $\pm$ 0.005 & 16.513 $\pm$ 0.007 & 16.273 $\pm$ 0.008 \\
3.4059 & 35 & \griz & 17.181 $\pm$ 0.007 & 16.771 $\pm$ 0.005 & 16.554 $\pm$ 0.008 & 16.309 $\pm$ 0.007 \\
4.1611 & 115 & \griz & 17.273 $\pm$ 0.005 & 16.869 $\pm$ 0.004 & 16.643 $\pm$ 0.005 & 16.418 $\pm$ 0.006 \\
4.4496 & 115 & \griz & 17.315 $\pm$ 0.005 & 16.911 $\pm$ 0.004 & 16.689 $\pm$ 0.005 & 16.452 $\pm$ 0.006 \\
4.6632 & 115 & \griz & 17.341 $\pm$ 0.005 & 16.944 $\pm$ 0.004 & 16.716 $\pm$ 0.005 & 16.488 $\pm$ 0.006 \\
5.6014 & 115 & \griz & 17.456 $\pm$ 0.005 & 17.067 $\pm$ 0.004 & 16.830 $\pm$ 0.005 & 16.601 $\pm$ 0.006 \\
5.8030 & 115 & \griz & 17.501 $\pm$ 0.006 & 17.094 $\pm$ 0.004 & 16.881 $\pm$ 0.006 & 16.637 $\pm$ 0.006 \\
6.0290 & 115 & \griz & 17.529 $\pm$ 0.005 & 17.140 $\pm$ 0.004 & 16.910 $\pm$ 0.006 & 16.675 $\pm$ 0.006 \\
6.2166 & 115 & \griz & 17.556 $\pm$ 0.006 & 17.163 $\pm$ 0.004 & 16.941 $\pm$ 0.005 & 16.700 $\pm$ 0.006 \\
6.5862 & 115 & \griz & 17.591 $\pm$ 0.006 & 17.197 $\pm$ 0.004 & 16.977 $\pm$ 0.006 & 16.746 $\pm$ 0.006 \\
6.7883 & 115 & \griz & 17.598 $\pm$ 0.006 & 17.219 $\pm$ 0.004 & 16.984 $\pm$ 0.006 & 16.757 $\pm$ 0.006 \\
6.9845 & 115 & \griz & 17.643 $\pm$ 0.006 & 17.254 $\pm$ 0.004 & 17.028 $\pm$ 0.006 & 16.788 $\pm$ 0.007 \\
7.2804 & 115 & \griz & 17.689 $\pm$ 0.006 & 17.300 $\pm$ 0.004 & 17.073 $\pm$ 0.006 & 16.839 $\pm$ 0.007 \\
7.4691 & 115 & \griz & 17.713 $\pm$ 0.006 & 17.323 $\pm$ 0.004 & 17.087 $\pm$ 0.006 & 16.853 $\pm$ 0.006 \\
7.6567 & 115 & \griz & 17.760 $\pm$ 0.005 & 17.342 $\pm$ 0.004 & 17.101 $\pm$ 0.004 & 16.902 $\pm$ 0.006 \\
7.8462 & 115 & \griz & 17.769 $\pm$ 0.005 & 17.355 $\pm$ 0.004 & 17.130 $\pm$ 0.005 & 16.918 $\pm$ 0.006 \\
8.0363 & 115 & \griz & 17.779 $\pm$ 0.004 & 17.371 $\pm$ 0.004 & 17.150 $\pm$ 0.005 & 16.918 $\pm$ 0.006 \\
8.2434 & 115 & \griz & 17.786 $\pm$ 0.006 & 17.397 $\pm$ 0.004 & 17.170 $\pm$ 0.007 & 16.928 $\pm$ 0.007 \\
8.4312 & 115 & \griz & 17.812 $\pm$ 0.005 & 17.423 $\pm$ 0.004 & 17.197 $\pm$ 0.005 & 16.989 $\pm$ 0.007 \\
8.6219 & 115 & \griz & 17.826 $\pm$ 0.004 & 17.432 $\pm$ 0.004 & 17.226 $\pm$ 0.004 & 17.005 $\pm$ 0.006 \\
8.8119 & 115 & \griz & 17.836 $\pm$ 0.006 & 17.461 $\pm$ 0.005 & 17.247 $\pm$ 0.005 & 17.010 $\pm$ 0.006 \\
9.0840 & 115 & \griz & 17.875 $\pm$ 0.006 & 17.498 $\pm$ 0.005 & 17.274 $\pm$ 0.008 & 17.043 $\pm$ 0.010 \\
9.2723 & 115 & \griz & 17.898 $\pm$ 0.006 & 17.527 $\pm$ 0.004 & 17.304 $\pm$ 0.006 & 17.086 $\pm$ 0.007 \\
9.4591 & 115 & \griz & 17.929 $\pm$ 0.005 & 17.534 $\pm$ 0.004 & 17.324 $\pm$ 0.006 & 17.123 $\pm$ 0.008 \\
9.6457 & 115 & \griz & 17.960 $\pm$ 0.005 & 17.570 $\pm$ 0.004 & 17.333 $\pm$ 0.005 & 17.116 $\pm$ 0.006 \\
9.8478 & 115 & \griz & 17.989 $\pm$ 0.006 & 17.599 $\pm$ 0.005 & 17.366 $\pm$ 0.007 & 17.132 $\pm$ 0.008 \\
10.037 & 115 & \griz & 18.013 $\pm$ 0.005 & 17.619 $\pm$ 0.004 & 17.401 $\pm$ 0.005 & 17.180 $\pm$ 0.006 \\
10.227 & 115 & \griz & 18.025 $\pm$ 0.005 & 17.636 $\pm$ 0.004 & 17.419 $\pm$ 0.005 & 17.206 $\pm$ 0.007 \\
10.420 & 115 & \griz & 18.046 $\pm$ 0.006 & 17.646 $\pm$ 0.005 & 17.408 $\pm$ 0.004 & 17.187 $\pm$ 0.008 \\
10.622 & 115 & \griz & 18.072 $\pm$ 0.006 & 17.674 $\pm$ 0.005 & 17.446 $\pm$ 0.007 & 17.203 $\pm$ 0.008 \\
10.811 & 115 & \griz & 18.091 $\pm$ 0.005 & 17.692 $\pm$ 0.004 & 17.467 $\pm$ 0.005 & 17.266 $\pm$ 0.006 \\
11.001 & 115 & \griz & 18.114 $\pm$ 0.005 & 17.718 $\pm$ 0.004 & 17.494 $\pm$ 0.005 & 17.281 $\pm$ 0.006 \\
11.194 & 115 & \griz & 18.142 $\pm$ 0.005 & 17.753 $\pm$ 0.007 & 17.523 $\pm$ 0.006 & 17.289 $\pm$ 0.006 \\
11.333 & 115 & \griz & 18.175 $\pm$ 0.006 & 17.776 $\pm$ 0.004 & 17.547 $\pm$ 0.007 & 17.302 $\pm$ 0.008 \\
11.574 & 115 & \griz & 18.199 $\pm$ 0.005 & 17.798 $\pm$ 0.005 & 17.567 $\pm$ 0.008 & 17.365 $\pm$ 0.006 \\
11.763 & 115 & \griz & 18.222 $\pm$ 0.005 & 17.822 $\pm$ 0.004 & 17.613 $\pm$ 0.008 & 17.391 $\pm$ 0.006 \\
11.957 & 115 & \griz & 18.258 $\pm$ 0.006 & 17.855 $\pm$ 0.009 & 17.633 $\pm$ 0.008 & 17.400 $\pm$ 0.007 \\
12.134 & 66 & \griz & 18.268 $\pm$ 0.011 & 17.877 $\pm$ 0.007 & 17.656 $\pm$ 0.010 & 17.395 $\pm$ 0.010 \\
12.270 & 66 & \griz & 18.300 $\pm$ 0.007 & 17.895 $\pm$ 0.005 & 17.671 $\pm$ 0.007 & 17.415 $\pm$ 0.010 \\
12.410 & 66 & \griz & 18.320 $\pm$ 0.011 & 17.915 $\pm$ 0.006 & 17.671 $\pm$ 0.009 & 17.462 $\pm$ 0.010 \\
12.552 & 66 & \griz & 18.348 $\pm$ 0.013 & 17.948 $\pm$ 0.008 & 17.714 $\pm$ 0.011 & 17.443 $\pm$ 0.013 \\
62.856 & 300 & R$^{(d)}$ & \multicolumn{4}{c}{20.49 $\pm$ 0.18} \\
64.523 & 4 x 300 & I$^{(d)}$ & \multicolumn{4}{c}{20.23 $\pm$ 0.09} \\
266.59 & 8 x 365 & \griz & 23.47 $\pm$ 0.06 & 23.02 $\pm$ 0.05 & 22.79 $\pm$ 0.07 & 22.59 $\pm$ 0.11 \\
353.11 & 8 x 365 & \griz & 24.09 $\pm$ 0.07 & 23.56 $\pm$ 0.06 & 23.28 $\pm$ 0.12 & 22.97 $\pm$ 0.15 \\
\hline
\end{tabular}

\noindent{$^{(a)}$ Not corrected for Galactic foreground reddening}  \\
  {$^{(b)}$ In the light curve fitting, a systematic error of 0.012 mag was added quadratically to the quoted statistical error}\\
  {$^{(c)}$ For the SED fitting, the aditional error of the absolute calibration of 0.05 mag was added}\\
  {$^{(d)}$ Calibrated using the GROND $r^\prime$ and $i^\prime$ field calibration, including a  ($r^\prime$-$i^\prime$) color term}\\
\end{table*}
}

%% file: tab01.tex
\onltab{2}{
\begin{table*}
\caption{JHK${_S}$ photometric data \label{JHKphot}}
\begin{tabular}{cccccc}
\hline
\noalign{\smallskip}
$T_{\rm mid} - T_{\rm 0} $ & Exposure [s] & Filter & \multicolumn{3}{c}{Brightness$^{(a)}$}  \\  
$[ks]$ & $[s]$ &  & \multicolumn{3}{c}{mag$_{\rm AB}^{(bc)}$}  \\ 
\hline
7.4943 & 12 x 10 & \JHK & 16.506 $\pm$ 0.009 & 16.215 $\pm$ 0.017 & 15.887 $\pm$ 0.019 \\
7.6818 & 12 x 10 & \JHK & 16.541 $\pm$ 0.008 & 16.240 $\pm$ 0.017 & 15.954 $\pm$ 0.019 \\
7.8710 & 12 x 10 & \JHK & 16.533 $\pm$ 0.008 & 16.251 $\pm$ 0.011 & 15.940 $\pm$ 0.013 \\
8.0611 & 12 x 10 & \JHK & 16.571 $\pm$ 0.008 & 16.264 $\pm$ 0.013 & 15.962 $\pm$ 0.014 \\
8.2685 & 12 x 10 & \JHK & 16.591 $\pm$ 0.009 & 16.267 $\pm$ 0.012 & 15.968 $\pm$ 0.014 \\
8.4560 & 12 x 10 & \JHK & 16.638 $\pm$ 0.008 & 16.329 $\pm$ 0.015 & 16.042 $\pm$ 0.016 \\
8.6469 & 12 x 10 & \JHK & 16.638 $\pm$ 0.008 & 16.309 $\pm$ 0.012 & 16.052 $\pm$ 0.014 \\
8.8370 & 12 x 10 & \JHK & 16.664 $\pm$ 0.009 & 16.346 $\pm$ 0.013 & 16.058 $\pm$ 0.015 \\
9.1091 & 12 x 10 & \JHK & 16.703 $\pm$ 0.009 & 16.378 $\pm$ 0.014 & 16.075 $\pm$ 0.016 \\
9.2966 & 12 x 10 & \JHK & 16.706 $\pm$ 0.008 & 16.397 $\pm$ 0.011 & 16.135 $\pm$ 0.013 \\
9.4841 & 12 x 10 & \JHK & 16.714 $\pm$ 0.008 & 16.397 $\pm$ 0.013 & 16.140 $\pm$ 0.015 \\
9.6707 & 12 x 10 & \JHK & 16.745 $\pm$ 0.009 & 16.437 $\pm$ 0.012 & 16.215 $\pm$ 0.014 \\
9.8729 & 12 x 10 & \JHK & 16.765 $\pm$ 0.009 & 16.455 $\pm$ 0.017 & 16.175 $\pm$ 0.018 \\
10.062 & 12 x 10 & \JHK & 16.825 $\pm$ 0.009 & 16.508 $\pm$ 0.015 & 16.252 $\pm$ 0.016 \\
10.252 & 12 x 10 & \JHK & 16.810 $\pm$ 0.008 & 16.533 $\pm$ 0.016 & 16.231 $\pm$ 0.017 \\
10.447 & 12 x 10 & \JHK & 16.842 $\pm$ 0.008 & 16.556 $\pm$ 0.017 & 16.258 $\pm$ 0.018 \\
10.646 & 12 x 10 & \JHK & 16.857 $\pm$ 0.009 & 16.599 $\pm$ 0.017 & 16.236 $\pm$ 0.018 \\
10.835 & 12 x 10 & \JHK & 16.908 $\pm$ 0.008 & 16.611 $\pm$ 0.017 & 16.310 $\pm$ 0.019 \\
11.025 & 12 x 10 & \JHK & 16.916 $\pm$ 0.008 & 16.620 $\pm$ 0.015 & 16.321 $\pm$ 0.016 \\
11.213 & 12 x 10 & \JHK & 16.949 $\pm$ 0.009 & 16.629 $\pm$ 0.014 & 16.334 $\pm$ 0.016 \\
11.415 & 12 x 10 & \JHK & 16.950 $\pm$ 0.009 & 16.658 $\pm$ 0.014 & 16.370 $\pm$ 0.016 \\
11.598 & 12 x 10 & \JHK & 16.989 $\pm$ 0.009 & 16.684 $\pm$ 0.011 & 16.411 $\pm$ 0.013 \\
11.788 & 12 x 10 & \JHK & 16.993 $\pm$ 0.008 & 16.737 $\pm$ 0.012 & 16.405 $\pm$ 0.014 \\
11.983 & 12 x 10 & \JHK & 17.071 $\pm$ 0.009 & 16.751 $\pm$ 0.013 & 16.414 $\pm$ 0.015 \\
12.140 & 6 x 10 & \JHK & 17.048 $\pm$ 0.011 & 16.770 $\pm$ 0.013 & 16.465 $\pm$ 0.014 \\
12.277 & 6 x 10 & \JHK & 17.079 $\pm$ 0.011 & 16.794 $\pm$ 0.014 & 16.453 $\pm$ 0.016 \\
12.417 & 6 x 10 & \JHK & 17.099 $\pm$ 0.010 & 16.788 $\pm$ 0.016 & 16.512 $\pm$ 0.017 \\
12.560 & 6 x 10 & \JHK & 17.141 $\pm$ 0.011 & 16.805 $\pm$ 0.016 & 16.458 $\pm$ 0.017 \\
12.707 & 6 x 10 & \JHK & 17.154 $\pm$ 0.010 & 16.796 $\pm$ 0.015 & 16.478 $\pm$ 0.016 \\
12.806 & 6 x 10 & \JHK & 17.144 $\pm$ 0.010 & 16.820 $\pm$ 0.014 & 16.543 $\pm$ 0.016 \\
12.904 & 6 x 10 & \JHK & 17.177 $\pm$ 0.011 & 16.808 $\pm$ 0.012 & 16.500 $\pm$ 0.014 \\
13.003 & 6 x 10 & \JHK & 17.174 $\pm$ 0.011 & 16.878 $\pm$ 0.024 & 16.553 $\pm$ 0.025 \\
13.116 & 6 x 10 & \JHK & 17.170 $\pm$ 0.011 & 16.863 $\pm$ 0.015 & 16.544 $\pm$ 0.016 \\
13.209 & 6 x 10 & \JHK & 17.226 $\pm$ 0.010 & 16.887 $\pm$ 0.017 & 16.555 $\pm$ 0.018 \\
13.308 & 6 x 10 & \JHK & 17.210 $\pm$ 0.010 & 16.856 $\pm$ 0.014 & 16.569 $\pm$ 0.016 \\
13.408 & 6 x 10 & \JHK & - - - & 16.883 $\pm$ 0.019 & 16.593 $\pm$ 0.020 \\
269.07 & 240 x 10 & \JHK & $>$ 22.47  & $>$ 21.97  & $>$ 21.224  \\
354.88 & 240 x 10 & \JHK & $>$ 22.29  & $>$ 22.04  & $>$ 21.082  \\
\hline
\end{tabular}

\noindent{$^{(a)}$ Not corrected for Galactic foreground reddening, but converted to AB magnitudes for consistency with Tab.~\ref{grizphot}}  \\
  {$^{(b)}$ In the light curve fitting, a systematic error of 0.02 mag was added quadratically to the quoted statistical error}\\
  {$^{(c)}$ For the SED fitting, the aditional error of the absolute calibration of 0.07 (J and H) and 0.09 (K) mag was added}
\end{table*}
}

%% file: tab1.tex
\onltab{3}{
\begin{table*}
\caption{Light curve fits \label{LCfits}}
\begin{tabular}{cccccccccc}
\hline
\noalign{\smallskip}
Bands &  $F_{\nu}(t)$ & $\alpha_{\rm r}^{(a)}$ & $s_{\rm 1}$ & $t_{\rm b,1}$ [s] & $\alpha_{d,1}^{(a)}$ & $s_{\rm 2}$ & $t_{\rm b,2}$ [s] & $\alpha_{\rm d,2}^{(a)}$ & $\chi^2$/d.o.f \\  
\hline
$g^\prime$ & TPL$^{(b)}$ & 1.20 $\pm$ 0.11 & 2.2 $\pm$ 0.5 & 1775 $\pm$ 62 & -0.64 $\pm$ 0.04 & 7.2 $\pm$ 1.7 & 9665 $\pm$ 170 & -1.58 $\pm$ 0.01 & 58 / 59 \\ 
$r^\prime$ & TPL$^{(b)}$ & 1.11 $\pm$ 0.07 & 2.6 $\pm$ 0.4 & 1816 $\pm$ 39 & -0.65 $\pm$ 0.03 & 6.7 $\pm$ 1.3 & 9767 $\pm$ 157 & -1.55 $\pm$ 0.01 & 49 / 60 \\ 
$i^\prime$ & TPL$^{(b)}$ & 1.10 $\pm$ 0.05 & 3.1 $\pm$ 0.5 & 1836 $\pm$ 37 & -0.63 $\pm$ 0.03 & 5.5 $\pm$ 1.2 & 9752 $\pm$ 185 & -1.56 $\pm$ 0.02 & 52 / 60 \\ 
$z^\prime$ & TPL$^{(b)}$ & 1.10 $\pm$ 0.06 & 3.4 $\pm$ 0.5 & 1835 $\pm$ 34 & -0.60 $\pm$ 0.04 & 4.2 $\pm$ 1.0 & 9795 $\pm$ 268 & -1.56 $\pm$ 0.03 & 61 / 59 \\ 
\hline
$JHK$ & DPL$^{(c)}$ & --- & --- & --- & -0.53 $\pm$ 0.15 & 5.7 $\pm$ 1.7 & 9542 $\pm$ 527 & -1.57 $\pm$ 0.15 & 84 / 99 \\ 
\hline
\gK & TPL$^{(b)}$ & 1.11 $\pm$ 0.03 & 2.9 $\pm$ 0.2 & 1829 $\pm$ 19 & -0.63 $\pm$ 0.02 & 5.7 $\pm$ 0.5 & 9763 $\pm$ 83 & -1.57 $\pm$ 0.01 & 425 / 362 \\ 
\hline
\gK +X-ray & TPL$^{(b)}$ & 1.11 $\pm$ 0.03 & 2.9 $\pm$ 0.2 & 1829 $\pm$ 19 & -0.63 $\pm$ 0.02 & 5.7 $\pm$ 0.5 & 9759 $\pm$ 82 & -1.57 $\pm$ 0.01 & 488 / 428 \\ 
\hline \\
\end{tabular}

\noindent{$^{(a)}$ Power law indices $\alpha$ of the segmented light curve, which are connected via breaks with smoothness s at break times t$_{\rm b}$  \\
  $^{(b)}$ Smoothly connected triple power law  \\
  $^{(c)}$ Smoothly connected double power law}
\end{table*}
}

%% file: tab2.tex
\onltab{4}{
\begin{table*}
\caption{SED fits \label{SEDfits}}
\begin{tabular}{cccccc}
\hline
\noalign{\smallskip}
Epoch &  Spectral index $\beta$ & N$_{\rm H}^{(a)}$ [10$^{22}\rm cm^2$] & $\chi^2$/d.o.f \\  
\hline
I & 1.00 $\pm$ 0.01 & 0.19 $\pm$ 0.09 & 36 / 36 \\ 
II & 0.99 $\pm$ 0.01 & 0.17 $\pm$ 0.10 & 15 / 15 \\ 
III & 1.01 $\pm$ 0.01 & 0.13 $^{+0.15}_{-0.13}$ & 18 / 19 \\ 
IV & 1.01 $\pm$ 0.01 & 0.53 $^{+1.30}_{-0.53}$ & 0.3 / 3 \\ 
\hline \\
\end{tabular}

\noindent{$^{(a)}$ Intrisic hydrogen column desity, in excess of the frozen Galactic foreground of $N_{\rm H}$=4.1$\times 10^{20}$~cm$^{-2}$} \\
\end{table*}
}

%% file: v2.bbl
\begin{thebibliography}{63}
\expandafter\ifx\csname natexlab\endcsname\relax\def\natexlab#1{#1}\fi

\bibitem[{{Abdo} {et~al.}(2009){Abdo}, {Ackermann}, \& {Arimoto}}]{abd09}
{Abdo}, A.~A., {Ackermann}, M., \& {Arimoto}, M. 2009, Science, 1688

\bibitem[{{Amati} {et~al.}(2002){Amati}, {Frontera}, {Tavani}, {in't Zand},
  {Antonelli}, {Costa}, {Feroci}, {Guidorzi}, {Heise}, {Masetti}, {Montanari},
  {Nicastro}, {Palazzi}, {Pian}, {Piro}, \& {Soffitta}}]{ama02}
{Amati}, L., {Frontera}, F., {Tavani}, M., {et~al.} 2002, \aap, 390, 81

\bibitem[{{Amati} {et~al.}(2008){Amati}, {Guidorzi}, {Frontera}, {Della Valle},
  {Finelli}, {Landi}, \& {Montanari}}]{ama08}
{Amati}, L., {Guidorzi}, C., {Frontera}, F., {et~al.} 2008, \mnras, 391, 577

\bibitem[{{Arnaud}(1996)}]{arn96}
{Arnaud}, K.~A. 1996, in ASPC Ser., Vol. 101, Astronomical Data Analysis
  Software and Systems V, ed. G.~H. {Jacoby} \& J.~{Barnes}, 17

\bibitem[{{Band} {et~al.}(1993){Band}, {Matteson}, {Ford}, {Schaefer},
  {Palmer}, {Teegarden}, {Cline}, {Briggs}, {Paciesas}, {Pendleton}, {Fishman},
  {Kouveliotou}, {Meegan}, {Wilson}, \& {Lestrade}}]{ban93}
{Band}, D., {Matteson}, J., {Ford}, L., {et~al.} 1993, \apj, 413, 281

\bibitem[{{Barthelmy} {et~al.}(2005)}]{bar05}
{Barthelmy}, S.~D. {et~al.} 2005, Space Science Reviews, 120, 143

\bibitem[{{Berger} {et~al.}(2003){Berger}, {Kulkarni}, {Pooley}, {Frail},
  {McIntyre}, {Wark}, {Sari}, {Soderberg}, {Fox}, {Yost}, \& {Price}}]{ber03}
{Berger}, E., {Kulkarni}, S.~R., {Pooley}, G., {et~al.} 2003, \nat, 426, 154

\bibitem[{{Burrows} {et~al.}(2005)}]{bur05}
{Burrows}, D.~N. {et~al.} 2005, Space Science Reviews, 120, 165

\bibitem[{{Della Valle} {et~al.}(2006){Della Valle}, {Chincarini}, {Panagia},
  {Tagliaferri}, {Malesani}, {Testa}, {Fugazza}, {Campana}, {Covino},
  {Mangano}, {Antonelli}, {D'Avanzo}, {Hurley}, {Mirabel}, {Pellizza},
  {Piranomonte}, \& {Stella}}]{del06}
{Della Valle}, M., {Chincarini}, G., {Panagia}, N., {et~al.} 2006, \nat, 444,
  1050

\bibitem[{{Eichler} \& {Granot}(2006)}]{eic06}
{Eichler}, D. \& {Granot}, J. 2006, \apjl, 641, L5

\bibitem[{{Evans} {et~al.}(2008){Evans}, {Beardmore}, {Page}, {Osborne},
  {O'Brien}, {Willingale}, {Starling}, {Burrows}, {Godet}, {Vetere}, {Racusin},
  {Goad}, {Wiersema}, {Angelini}, {Capalbi}, {Chincarini}, {Gehrels}, {Kennea},
  {Margutti}, {Morris}, {Mountford}, {Pagani}, {Perri}, {Romano}, \&
  {Tanvir}}]{eva09}
{Evans}, P.~A., {Beardmore}, A.~P., {Page}, K.~L., {et~al.} 2008,
  ArXiv:0812.3662

\bibitem[{{Evans} {et~al.}(2007)}]{eva07}
{Evans}, P.~A. {et~al.} 2007, \aap, 469, 379

\bibitem[{{Ferrero} {et~al.}(2009)}]{fer09}
{Ferrero}, P. {et~al.} 2009, \aap, 497, 729

\bibitem[{{Fynbo} {et~al.}(2009){Fynbo}, {Jakobsson}, {Prochaska}, {Malesani},
  {Ledoux}, {de Ugarte Postigo}, {Nardini}, {Vreeswijk}, {Hjorth}, {Sollerman},
  {Chen}, {Thoene}, {Bjoernsson}, {Bloom}, {Castro-Tirado}, {Christensen}, {De
  Cia}, {Gorosabel}, {Jaunsen}, {Jensen}, {Levan}, {Maund}, {Masetti},
  {Milvang-Jensen}, {Palazzi}, {Perley}, {Pian}, {Rol}, {Schady}, {Starling},
  {Tanvir}, {Watson}, {Wiersema}, {Xu}, {Augusteijn}, {Grundahl}, {Telting}, \&
  {Quirion}}]{fyn09}
{Fynbo}, J.~P.~U., {Jakobsson}, P., {Prochaska}, J.~X., {et~al.} 2009,
  ArXiv:0907.3449

\bibitem[{{Fynbo} {et~al.}(2004){Fynbo}, {Sollerman}, {Hjorth}, {Grundahl},
  {Gorosabel}, {Weidinger}, {M{\o}ller}, {Jensen}, {Vreeswijk}, {Fransson},
  {Ramirez-Ruiz}, {Jakobsson}, {J{\o}rgensen}, {Vinter}, {Andersen}, {Castro
  Cer{\'o}n}, {Castro-Tirado}, {Fruchter}, {Greiner}, {Kouveliotou}, {Levan},
  {Klose}, {Masetti}, {Pedersen}, {Palazzi}, {Pian}, {Rhoads}, {Rol},
  {Sekiguchi}, {Tanvir}, {Tristram}, {de Ugarte Postigo}, {Wijers}, \& {van den
  Heuvel}}]{fyn04}
{Fynbo}, J.~P.~U., {Sollerman}, J., {Hjorth}, J., {et~al.} 2004, \apj, 609, 962

\bibitem[{{Gehrels} {et~al.}(2004)}]{geh04}
{Gehrels}, N. {et~al.} 2004, \apj, 611, 1005

\bibitem[{{Granot} {et~al.}(2006){Granot}, {K{\"o}nigl}, \& {Piran}}]{gran06}
{Granot}, J., {K{\"o}nigl}, A., \& {Piran}, T. 2006, \mnras, 370, 1946

\bibitem[{{Granot} \& {Kumar}(2003)}]{gra03a}
{Granot}, J. \& {Kumar}, P. 2003, \apj, 591, 1086

\bibitem[{{Granot} {et~al.}(2002){Granot}, {Panaitescu}, {Kumar}, \&
  {Woosley}}]{gra02}
{Granot}, J., {Panaitescu}, A., {Kumar}, P., \& {Woosley}, S.~E. 2002, \apjl,
  570, L61

\bibitem[{{Greiner} {et~al.}(2008){Greiner}, {Bornemann}, {Clemens}, {Deuter},
  {Hasinger}, {Honsberg}, {Huber}, {Huber}, {Krauss}, {Kr{\"u}hler},
  {K{\"u}pc{\"u} Yolda{\c s}}, {Mayer-Hasselwander}, {Mican}, {Primak},
  {Schrey}, {Steiner}, {Szokoly}, {Th{\"o}ne}, {Yolda{\c s}}, {Klose}, {Laux},
  \& {Winkler}}]{gre08}
{Greiner}, J., {Bornemann}, W., {Clemens}, C., {et~al.} 2008, \pasp, 120, 405

\bibitem[{{Greiner} {et~al.}(2009{\natexlab{a}}){Greiner}, {Clemens},
  {Kr{\"u}hler}, {Kienlin}, {Rau}, {Sari}, {Fox}, {Kawai}, {Afonso}, {Ajello},
  {Berger}, {Cenko}, {Cucchiara}, {Filgas}, {Klose}, {Kuepue Yoldas}, {Lichti},
  {Loew}, {McBreen}, {Nagayama}, {Rossi}, {Sato}, {Szokoly}, {Yoldas}, \&
  {Zhang}}]{gre09}
{Greiner}, J., {Clemens}, C., {Kr{\"u}hler}, T., {et~al.} 2009{\natexlab{a}},
  \aap, 498, 89

\bibitem[{{Greiner} {et~al.}(2009{\natexlab{b}}){Greiner}, {Kr{\"u}hler},
  {McBreen}, {Ajello}, {Giannios}, {Schwarz}, {Savaglio}, {Yolda{\c s}},
  {Clemens}, {Stefanescu}, {Sala}, {Bertoldi}, {Szokoly}, \& {Klose}}]{gre09a}
{Greiner}, J., {Kr{\"u}hler}, T., {McBreen}, S., {et~al.} 2009{\natexlab{b}},
  \apj, 693, 1912

\bibitem[{{Guidorzi} {et~al.}(2009){Guidorzi}, {Clemens}, {Kobayashi},
  {Granot}, {Melandri}, {D'Avanzo}, {Kuin}, {Klotz}, {Fynbo}, {Covino},
  {Greiner}, {Malesani}, {Mao}, {Mundell}, {Steele}, {Jakobsson}, {Margutti},
  {Bersier}, {Campana}, {Chincarini}, {D'Elia}, {Fugazza}, {Genet}, {Gomboc},
  {Kr{\"u}hler}, {K{\"u}pc{\"u} Yolda{\c s}}, {Moretti}, {Mottram}, {O'Brien},
  {Smith}, {Szokoly}, {Tagliaferri}, {Tanvir}, \& {Gehrels}}]{gui09}
{Guidorzi}, C., {Clemens}, C., {Kobayashi}, S., {et~al.} 2009, \aap, 499, 439

\bibitem[{{Heise} {et~al.}(2001){Heise}, {in't Zand}, {Kippen}, \&
  {Woods}}]{hei01}
{Heise}, J., {in't Zand}, J., {Kippen}, R.~M., \& {Woods}, P.~M. 2001, in
  Gamma-ray Bursts in the Afterglow Era, ed. E.~{Costa}, F.~{Frontera}, \&
  J.~{Hjorth}, 16

\bibitem[{{Kalberla} {et~al.}(2005){Kalberla}, {Burton}, {Hartmann}, {Arnal},
  {Bajaja}, {Morras}, \& {P{\"o}ppel}}]{kal05}
{Kalberla}, P.~M.~W., {Burton}, W.~B., {Hartmann}, D., {et~al.} 2005, \aap,
  440, 775

\bibitem[{{Kippen} {et~al.}(2003){Kippen}, {Woods}, {Heise}, {in't Zand},
  {Briggs}, \& {Preece}}]{kip03}
{Kippen}, R.~M., {Woods}, P.~M., {Heise}, J., {et~al.} 2003, in AIPC, Vol. 662,
  Gamma-Ray Burst and Afterglow Astronomy 2001, ed. G.~R. {Ricker} \& R.~K.
  {Vanderspek}, 244

\bibitem[{{Klotz} {et~al.}(2009){Klotz}, {Bo{\"e}r}, {Atteia}, \&
  {Gendre}}]{klo09}
{Klotz}, A., {Bo{\"e}r}, M., {Atteia}, J.~L., \& {Gendre}, B. 2009, \aj, 137,
  4100

\bibitem[{{Kr{\"u}hler} {et~al.}(2009){Kr{\"u}hler}, {Greiner}, {McBreen},
  {Klose}, {Rossi}, {Afonso}, {Clemens}, {Filgas}, {K{\"u}pc{\"u} Yoldas},
  {Szokoly}, \& {Yoldas}}]{tk09}
{Kr{\"u}hler}, T., {Greiner}, J., {McBreen}, S., {et~al.} 2009, \apj, 697, 758

\bibitem[{{Kr{\"u}hler} {et~al.}(2008){Kr{\"u}hler}, {K{\"u}pc{\"u} Yolda{\c
  s}}, {Greiner}, {Clemens}, {McBreen}, {Primak}, {Savaglio}, {Yolda{\c s}},
  {Szokoly}, \& {Klose}}]{tk08}
{Kr{\"u}hler}, T., {K{\"u}pc{\"u} Yolda{\c s}}, A., {Greiner}, J., {et~al.}
  2008, \apj, 685, 376

\bibitem[{{Kumar} \& {Granot}(2003)}]{kum03}
{Kumar}, P. \& {Granot}, J. 2003, \apj, 591, 1075

\bibitem[{{Lamb} {et~al.}(2005){Lamb}, {Donaghy}, \& {Graziani}}]{lam05}
{Lamb}, D.~Q., {Donaghy}, T.~Q., \& {Graziani}, C. 2005, \apj, 620, 355

\bibitem[{{Landsman} \& {Sbarufatti}(2008)}]{lan08}
{Landsman}, W.~B. \& {Sbarufatti}, B. 2008, GCN, 7965

\bibitem[{{Liang} {et~al.}(2008){Liang}, {Racusin}, {Zhang}, {Zhang}, \&
  {Burrows}}]{lia08}
{Liang}, E.-W., {Racusin}, J.~L., {Zhang}, B., {Zhang}, B.-B., \& {Burrows},
  D.~N. 2008, \apj, 675, 528

\bibitem[{{M{\'e}sz{\'a}ros} {et~al.}(1998){M{\'e}sz{\'a}ros}, {Rees}, \&
  {Wijers}}]{mes98}
{M{\'e}sz{\'a}ros}, P., {Rees}, M.~J., \& {Wijers}, R.~A.~M.~J. 1998, \apj,
  499, 301

\bibitem[{{Molinari} {et~al.}(2007){Molinari}, {Vergani}, {Malesani}, {Covino},
  {D'Avanzo}, {Chincarini}, {Zerbi}, {Antonelli}, {Conconi}, {Testa}, {Tosti},
  {Vitali}, {D'Alessio}, {Malaspina}, {Nicastro}, {Palazzi}, {Guetta},
  {Campana}, {Goldoni}, {Masetti}, {Meurs}, {Monfardini}, {Norci}, {Pian},
  {Piranomonte}, {Rizzuto}, {Stefanon}, {Stella}, {Tagliaferri}, {Ward},
  {Ihle}, {Gonzalez}, {Pizarro}, {Sinclaire}, \& {Valenzuela}}]{mol07}
{Molinari}, E., {Vergani}, S.~D., {Malesani}, D., {et~al.} 2007, \aap, 469, L13

\bibitem[{{Nousek} {et~al.}(2006){Nousek}, {Kouveliotou}, {Grupe}, {Page},
  {Granot}, {Ramirez-Ruiz}, {Patel}, {Burrows}, {Mangano}, {Barthelmy},
  {Beardmore}, {Campana}, {Capalbi}, {Chincarini}, {Cusumano}, {Falcone},
  {Gehrels}, {Giommi}, {Goad}, {Godet}, {Hurkett}, {Kennea}, {Moretti},
  {O'Brien}, {Osborne}, {Romano}, {Tagliaferri}, \& {Wells}}]{nou06}
{Nousek}, J.~A., {Kouveliotou}, C., {Grupe}, D., {et~al.} 2006, \apj, 642, 389

\bibitem[{{Oates} {et~al.}(2009){Oates}, {Page}, {Schady}, {de Pasquale},
  {Koch}, {Breeveld}, {Brown}, {Chester}, {Holland}, {Hoversten}, {Kuin},
  {Marshall}, {Roming}, {Still}, {vanden Berk}, {Zane}, \& {Nousek}}]{oat09}
{Oates}, S.~R., {Page}, M.~J., {Schady}, P., {et~al.} 2009, \mnras, 395, 490

\bibitem[{{Panaitescu} {et~al.}(1998){Panaitescu}, {M{\'e}sz{\'a}ros}, \&
  {Rees}}]{pan98}
{Panaitescu}, A., {M{\'e}sz{\'a}ros}, P., \& {Rees}, M.~J. 1998, \apj, 503, 314

\bibitem[{{Panaitescu} \& {Vestrand}(2008)}]{pan08}
{Panaitescu}, A. \& {Vestrand}, W.~T. 2008, \mnras, 387, 497

\bibitem[{{Peng} {et~al.}(2005){Peng}, {K{\"o}nigl}, \& {Granot}}]{pen05}
{Peng}, F., {K{\"o}nigl}, A., \& {Granot}, J. 2005, \apj, 626, 966

\bibitem[{{Perley} {et~al.}(2008){Perley}, {Chornock}, \& {Bloom}}]{per08b}
{Perley}, D.~A., {Chornock}, R., \& {Bloom}, J.~S. 2008, GCN, 7962

\bibitem[{{Piran}(2005)}]{pir05}
{Piran}, T. 2005, Reviews of Modern Physics, 76, 1143

\bibitem[{{Preece} {et~al.}(2000){Preece}, {Briggs}, {Mallozzi}, {Pendleton},
  {Paciesas}, \& {Band}}]{pre00}
{Preece}, R.~D., {Briggs}, M.~S., {Mallozzi}, R.~S., {et~al.} 2000, \apjs, 126,
  19

\bibitem[{{Racusin} {et~al.}(2008){Racusin}, {Karpov}, {Sokolowski}, {Granot},
  {Wu}, {Pal'Shin}, {Covino}, {van der Horst}, {Oates}, {Schady}, {Smith},
  {Cummings}, {Starling}, {Piotrowski}, {Zhang}, {Evans}, {Holland}, {Malek},
  {Page}, {Vetere}, {Margutti}, {Guidorzi}, {Kamble}, {Curran}, {Beardmore},
  {Kouveliotou}, {Mankiewicz}, {Melandri}, {O'Brien}, {Page}, {Piran},
  {Tanvir}, {Wrochna}, {Aptekar}, {Barthelmy}, {Bartolini}, {Beskin}, {Bondar},
  {Bremer}, {Campana}, {Castro-Tirado}, {Cucchiara}, {Cwiok}, {D'Avanzo},
  {D'Elia}, {Della Valle}, {de Ugarte Postigo}, {Dominik}, {Falcone}, {Fiore},
  {Fox}, {Frederiks}, {Fruchter}, {Fugazza}, {Garrett}, {Gehrels},
  {Golenetskii}, {Gomboc}, {Gorosabel}, {Greco}, {Guarnieri}, {Immler},
  {Jelinek}, {Kasprowicz}, {La Parola}, {Levan}, {Mangano}, {Mazets},
  {Molinari}, {Moretti}, {Nawrocki}, {Oleynik}, {Osborne}, {Pagani}, {Pandey},
  {Paragi}, {Perri}, {Piccioni}, {Ramirez-Ruiz}, {Roming}, {Steele}, {Strom},
  {Testa}, {Tosti}, {Ulanov}, {Wiersema}, {Wijers}, {Winters}, {Zarnecki},
  {Zerbi}, {M{\'e}sz{\'a}ros}, {Chincarini}, \& {Burrows}}]{rac08}
{Racusin}, J.~L., {Karpov}, S.~V., {Sokolowski}, M., {et~al.} 2008, \nat, 455,
  183

\bibitem[{{Rees} \& {Meszaros}(1998)}]{ree98}
{Rees}, M.~J. \& {Meszaros}, P. 1998, \apjl, 496, L1+

\bibitem[{{Rhoads}(1999)}]{rho99}
{Rhoads}, J.~E. 1999, \apj, 525, 737

\bibitem[{{Roming} {et~al.}(2005)}]{rom05}
{Roming}, P.~W.~A. {et~al.} 2005, Space Science Reviews, 120, 95

\bibitem[{{Rossi} {et~al.}(2002){Rossi}, {Lazzati}, \& {Rees}}]{ros02}
{Rossi}, E., {Lazzati}, D., \& {Rees}, M.~J. 2002, \mnras, 332, 945

\bibitem[{{Rykoff} {et~al.}(2009){Rykoff}, {Aharonian}, {Akerlof}, {Ashley},
  {Barthelmy}, {Flewelling}, {Gehrels}, {Gogus}, {Guver}, {Kiziloglu}, {Krimm},
  {McKay}, {Ozel}, {Phillips}, {Quimby}, {Rowell}, {Rujopakarn}, {Schaefer},
  {Smith}, {Vestrand}, {Wheeler}, {Wren}, {Yuan}, \& {Yost}}]{ryk09}
{Rykoff}, E.~S., {Aharonian}, F., {Akerlof}, C.~W., {et~al.} 2009, ArXiv:
  0904.0261

\bibitem[{{Sakamoto} {et~al.}(2008){Sakamoto}, {Hullinger}, {Sato}, {Yamazaki},
  {Barbier}, {Barthelmy}, {Cummings}, {Fenimore}, {Gehrels}, {Krimm}, {Lamb},
  {Markwardt}, {Osborne}, {Palmer}, {Parsons}, {Stamatikos}, \&
  {Tueller}}]{sak08}
{Sakamoto}, T., {Hullinger}, D., {Sato}, G., {et~al.} 2008, \apj, 679, 570

\bibitem[{{Sakamoto} {et~al.}(2005){Sakamoto}, {Lamb}, {Kawai}, {Yoshida},
  {Graziani}, {Fenimore}, {Donaghy}, {Matsuoka}, {Suzuki}, {Ricker}, {Atteia},
  {Shirasaki}, {Tamagawa}, {Torii}, {Galassi}, {Doty}, {Vanderspek}, {Crew},
  {Villasenor}, {Butler}, {Prigozhin}, {Jernigan}, {Barraud}, {Boer},
  {Dezalay}, {Olive}, {Hurley}, {Levine}, {Monnelly}, {Martel}, {Morgan},
  {Woosley}, {Cline}, {Braga}, {Manchanda}, {Pizzichini}, {Takagishi}, \&
  {Yamauchi}}]{sak05}
{Sakamoto}, T., {Lamb}, D.~Q., {Kawai}, N., {et~al.} 2005, \apj, 629, 311

\bibitem[{{Sakamoto} {et~al.}(2009){Sakamoto}, {Sato}, {Barbier}, {Barthelmy},
  {Cummings}, {Fenimore}, {Gehrels}, {Hullinger}, {Krimm}, {Lamb}, {Markwardt},
  {Palmer}, {Parsons}, {Stamatikos}, {Tueller}, \& {Ukwatta}}]{sak09}
{Sakamoto}, T., {Sato}, G., {Barbier}, L., {et~al.} 2009, \apj, 693, 922

\bibitem[{{Sari} {et~al.}(1999){Sari}, {Piran}, \& {Halpern}}]{sar99}
{Sari}, R., {Piran}, T., \& {Halpern}, J.~P. 1999, \apjl, 519, L17

\bibitem[{{Sbarufatti} {et~al.}(2008){Sbarufatti}, {Baumgartner}, {Evans},
  {Guidorzi}, {Hoversten}, {La Parola}, {Mangano}, {Markwardt}, {Page},
  {Romano}, \& {Ukwatta}}]{sba08a}
{Sbarufatti}, B., {Baumgartner}, W.~H., {Evans}, P.~A., {et~al.} 2008, GCN,
  7957

\bibitem[{{Schlegel} {et~al.}(1998){Schlegel}, {Finkbeiner}, \&
  {Davis}}]{schl98}
{Schlegel}, D.~J., {Finkbeiner}, D.~P., \& {Davis}, M. 1998, \apj, 500, 525

\bibitem[{{Schulze} {et~al.}(2008){Schulze}, {Kann}, {Rossi}, {Gonsalves},
  {Hoegner}, \& {Stecklum}}]{sch08}
{Schulze}, S., {Kann}, D.~A., {Rossi}, A., {et~al.} 2008, GCN, 7972

\bibitem[{{Tody}(1993)}]{tod93}
{Tody}, D. 1993, in ASPC Ser., Vol.~52, Astronomical Data Analysis Software and
  Systems II, ed. R.~J. {Hanisch}, R.~J.~V. {Brissenden}, \& J.~{Barnes}, 173

\bibitem[{{Tueller} {et~al.}(2008){Tueller}, {Barthelmy}, {Baumgartner},
  {Cummings}, {Fenimore}, {Gehrels}, {Krimm}, {Markwardt}, {McLean}, {Palmer},
  {Parsons}, {Sakamoto}, {Sato}, {Stamatikos}, \& {Ukwatta}}]{tue08}
{Tueller}, J., {Barthelmy}, S.~D., {Baumgartner}, W., {et~al.} 2008, GCN, 7969

\bibitem[{{Woods} \& {Loeb}(1999)}]{woo99}
{Woods}, E. \& {Loeb}, A. 1999, \apj, 523, 187

\bibitem[{{Yamazaki} {et~al.}(2002){Yamazaki}, {Ioka}, \& {Nakamura}}]{yam02}
{Yamazaki}, R., {Ioka}, K., \& {Nakamura}, T. 2002, \apjl, 571, L31

\bibitem[{{Zhang} {et~al.}(2006){Zhang}, {Fan}, {Dyks}, {Kobayashi},
  {M{\'e}sz{\'a}ros}, {Burrows}, {Nousek}, \& {Gehrels}}]{zha06}
{Zhang}, B., {Fan}, Y.~Z., {Dyks}, J., {et~al.} 2006, \apj, 642, 354

\bibitem[{{Zhang} \& {M{\'e}sz{\'a}ros}(2002)}]{zha02b}
{Zhang}, B. \& {M{\'e}sz{\'a}ros}, P. 2002, \apj, 571, 876

\bibitem[{{Zhang} {et~al.}(2003){Zhang}, {Woosley}, \& {MacFadyen}}]{zhang03}
{Zhang}, W., {Woosley}, S.~E., \& {MacFadyen}, A.~I. 2003, \apj, 586, 356

\end{thebibliography}
